# Enhanced spin coherence at the sweet spot of a self-assembled quantum dot molecule

Kha X. Tran,[1] Allan S. Bracker,[2] Michael K. Yakes,[2†] Joel Q. Grim,[2], and Samuel G. Carter[2,*]
[1] NRC Research Associate at the Naval Research Laboratory, 4555 Overlook Ave. SW, Washington, DC 20375, USA
[2] Naval Research Laboratory, 4555 Overlook Ave. SW, Washington, DC 20375, USA

[†]Present address: Air Force Office of Scientific Research, Arlington, VA 22203, USA.
*sam.carter@nrl.navy.mil

*Abstract:*

A pair of coupled dots with one electron in each dot can provide improvements in spin coherence, particularly at an electrical bias called the "sweet spot", but few measurements have been performed on self-assembled dots in this regime. Here, we directly measure the $T_2^*$ coherence time of the singlet-triplet states in this system as a function of bias and magnetic field, obtaining a maximum $T_2^*$ of 60 ns, more than an order of magnitude higher than an electron spin in a single quantum dot. Our results uncover two main dephasing mechanisms: electrical noise away from the sweet spot, and a magnetic field dependent interaction with nuclear spins due to a difference in g-factors.

*Main text:*

The electron spin in an InGaAs self-assembled quantum dot (QD) is a promising candidate for quantum applications due to bright, spin-dependent optical transitions, fast optical control, integration, and potential scalability [1]. QDs are readily grown by molecular beam epitaxy and can be integrated into nanophotonic structures containing both electronic [2] and photonic circuit elements [3,4]. However, the electron spin in a QD is prone to decoherence, primarily due to interactions with the nuclear spin bath [5–7]. Without advanced spectroscopy techniques [8–13] to limit the effect of these interactions, the coherence time of a QD single spin is very short, i.e. a few nanoseconds [8,14,15], which severely limits its use.

One way to extend the coherence time in the semiconductor QD is to utilize the two electron spins in the coupled quantum dot system, which consists of the top and bottom dot separated by a tunneling barrier. The electrons hybridize and form the bonding and anti-bonding states, hence the name quantum dot molecule (QDM) [16–21]. The two electrons interact strongly with each other via the electron-electron exchange, which splits the energy degeneracy between the singlet (S) and triplet ($T_0$, T+, T-) ground states. The $m_s = 0$ S and $T_0$ states are in principle insensitive to fluctuating magnetic fields from nuclear spins [22], but the exchange splitting does depend on the electric field, making it sensitive to electric field fluctuations [23]. As a function of electrical bias, the QDM features a sweet spot which is the location of minimum exchange interaction where the S-$T_0$ subspace is first order immune to both electric and magnetic field fluctuations [22,24,25]. Indeed, operating at the sweet spot has proved to be a viable strategy to extend the coherence time of the electron spin. In the optically active QDM, $T_2^*$ is shown to have the upper limit of about 200 ns which is measured indirectly using the coherent population trapping technique [22].

In this letter, by exploiting the favorable characteristics of the QDM at the sweet spot, we demonstrate a $T_2^*$ coherence time of 60 ns, the longest value for an optically active QD that has been measured directly. We uncover two main dephasing mechanisms by examining $T_2^*$ as a function of electrical bias as well as the external magnetic field. The dephasing rate from electric field fluctuations is proportional to the derivative of the exchange interaction with respect to the electrical bias. This suggests that the slow electrical noise gives rise to fluctuations in the exchange splitting and limits the coherence away from the sweet spot [23]. Near the sweet spot, the coherence time is limited by the hyperfine interaction with nuclear spins, which has a strong effect at very low magnetic fields, due to mixing of triplet states, and also a strong effect at high magnetic fields, due to an effective magnetic field gradient [26] stemming from the g-factor difference between the two dots. This gives rise to a maximum $T_2^*$ at a modest magnetic field of 50 mT. To the best of our knowledge, this is the first report that measures $T_2^*$ directly and shows huge enhancement at the sweet spot in the GaAs QDM.

Our device is grown epitaxially and is embedded in a Bragg cavity with the n-i-p-i-p diode structure that allows the QDM to be deterministically charged with two electrons [27,28]. The sample is positioned at the center of a superconducting magnet ∼3 K which exerts an external magnetic field in the Faraday geometry. With one electron residing in each dot, the spin singlet (S) and spin triplet ($T_0$, T+, T-) ground states together with the trion excited states (R±, R± ±) form an eight-level system as shown in Fig. 1(a). To achieve electron tunneling, our QDM sample is designed so that the bottom dot transition energy is blue-shifted compared to that of the top dot [17,28]. Thus, we denote the blue (red) arrow as the bottom (top) dot spin [Fig. 1(a)]. The single (double) arrow refers to electron (hole). Due to the optical selection rules, there are six allowed transitions with two pairs of them being nearly degenerate. We thus label and color code the transitions from 1 to 4 in the order of increasing energy.

Fig. 1(b) depicts the bias map of the resonance fluorescence (RF) spectra taken at 1 T magnetic field showing the corresponding transitions 1 to 4. The two electron charge state is found to be stable for a bias range of 0.7 to 0.76 V. The RF signal is suppressed in the middle of the stable bias range due to optical spin pumping [29–32] where the spin is excited out of one spin state and shelved in the other after a few recombination cycles. The transition 1 and 3 are still relatively bright in the optical pumping region due to the fact that T-↔R-- and T+↔R++ are cycling transitions [33]. The exchange interaction i.e. the energy splitting between S and $T_0$ states, denoted by $J$ can be measured directly by taking the energy difference between transition 2 and 1 or transition 4 and 3. The change in relative energy levels of the electron residing in each dot with respect to the electric field leads to a $J$ that is dependent on the gate voltage. In this sample, we find the minimum $J$ (~ 45 μeV) (the sweet spot) to be at the center of the optical pumping region (0.725 V).

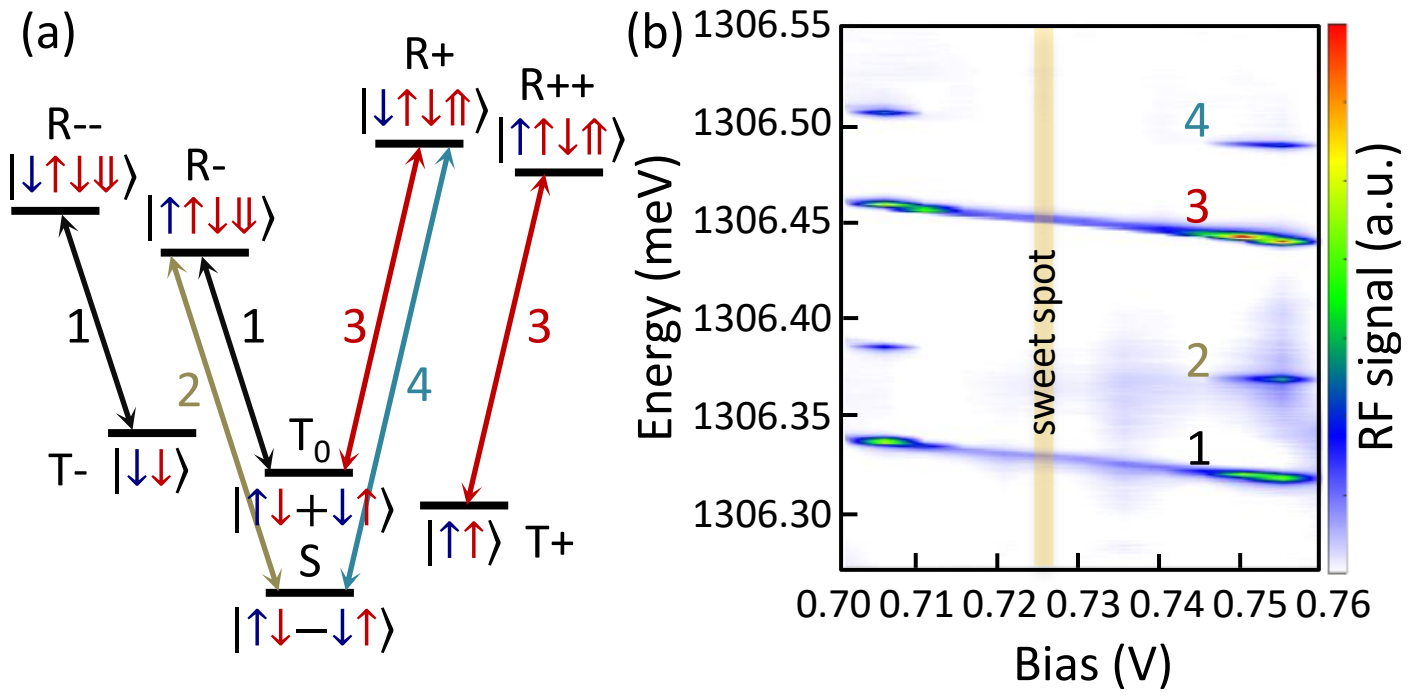

FIG. 1 (color online). (a) Detailed energy-level diagram of the two-electron charged QDM in the Faraday external magnetic field with numbering scheme depicting individual transitions. The spin configuration for each state is also indicated. Here, blue (red) arrow refers to the bottom (top) dot. Single (double) arrow refers to electron (hole). (b) Measured resonance fluorescence spectra as functions of bias. The transitions are identified with numbering and colors that match part (a). The sweet spot is at the center of the optical pumping region (about 0.725 V bias).

To perform electron spin control, we utilize the Λ system $T_0$↔R-↔S as shown in Fig. 2(a). The linearly polarized initialization laser (70 ns duration) resonant with $T_0$↔R- and T-↔R-- transitions pumps the spin into the S state [34]. Then, a circularly polarized rotation laser detuned from the excited state ($\Delta_L$~250 GHz) coherently rotates the spin between S and $T_0$ states, depending on the pulse duration and peak power. Finally, the $T_0$ population is detected via driving the $T_0$↔R- transition and monitoring the emission from the Raman transition S↔R-. We modulate the rotation laser by a microwave source via an electro-optic modulator (EOM) biased to zero transmission, which creates two side bands in the laser spectrum [12,35]. The side bands energy spacing is twice the modulation frequency, and thus can be varied to detune from the $ST_0$ exchange splitting a variable amount δ. We note that the energy levels in Fig. 2(a) are not drawn to scale. In reality, $\Delta_L$~250 GHz is about three order of magnitude larger than δ which is only tens to hundreds of MHz. Fig. 2(b) shows the evolution of the $T_0$ population which is normalized to 1 for increasing rotation pulse duration at four different rotation peak powers. Here, the microwave modulation is slightly detuned from the spin resonance. Thus, the linear fit to the Rabi frequency vs. peak power yields ~28.5 MHz y-intercept [Fig. 2(c)]. Low laser peak power and the relatively large laser detuning $\Delta_L$~250 GHz ensures that we are largely in the adiabatic limit where the excited states population is negligible [12,36].

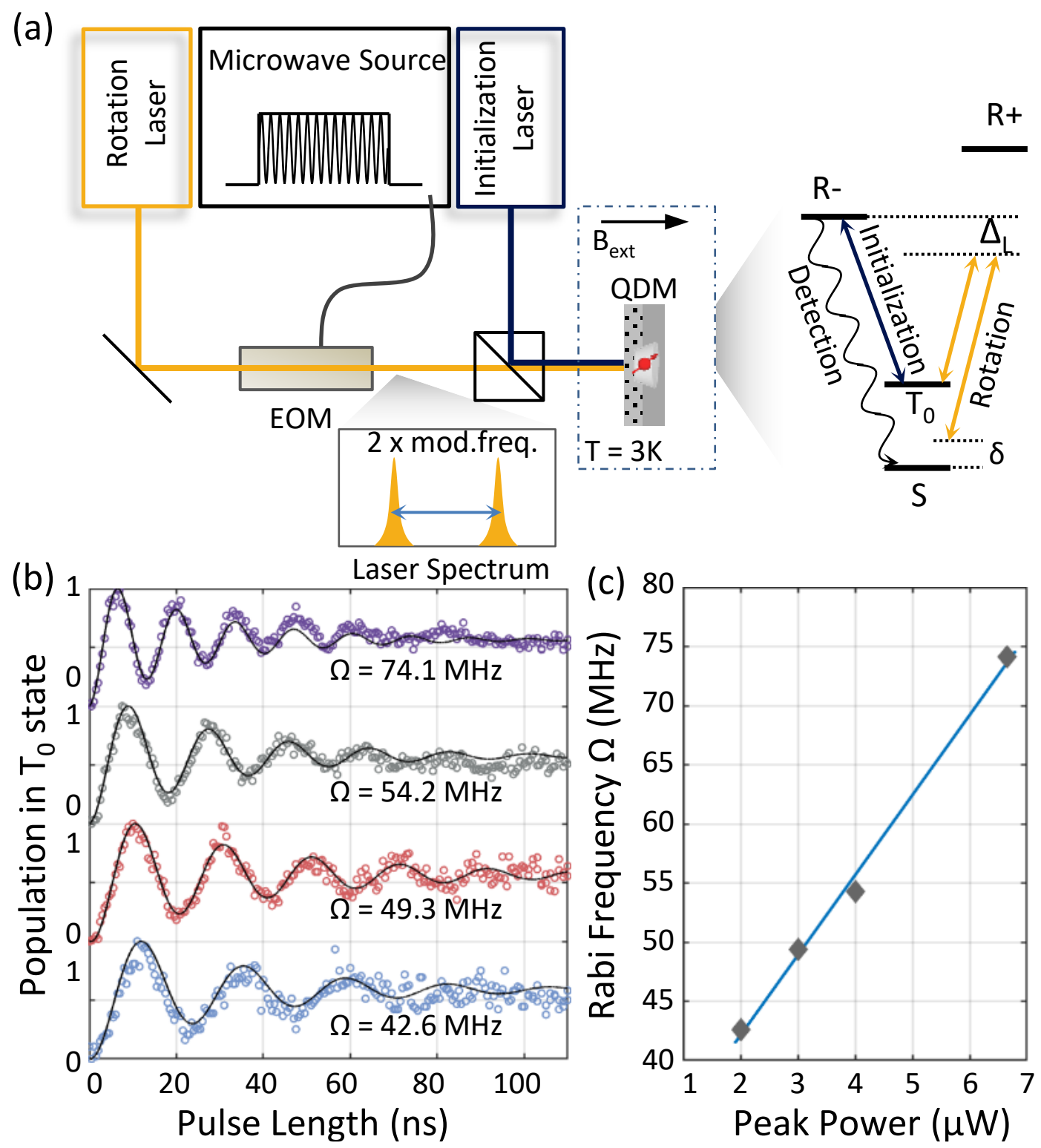

FIG. 2 (color online). (a) Experimental schematic: Intensity modulation of the rotation laser by a microwave source via the electro-optical modulator (EOM) creates two laser side bands whose energy difference is twice the modulation frequency. The rotation laser has a single-photon detuning of $\Delta_L \approx 250$ GHz from the excited state, and two-photon detuning of $\delta$ from the S-$T_0$ resonance. The initialization laser is resonant with $T_0 \leftrightarrow$R- transition. The $T_0$ state population is then detected via Raman scattering S$\leftrightarrow$R-. (b) Rabi oscillations between S and $T_0$ spin states at four different Rabi frequencies. (c) The dependence of the Rabi frequency on the peak power of the rotation laser.

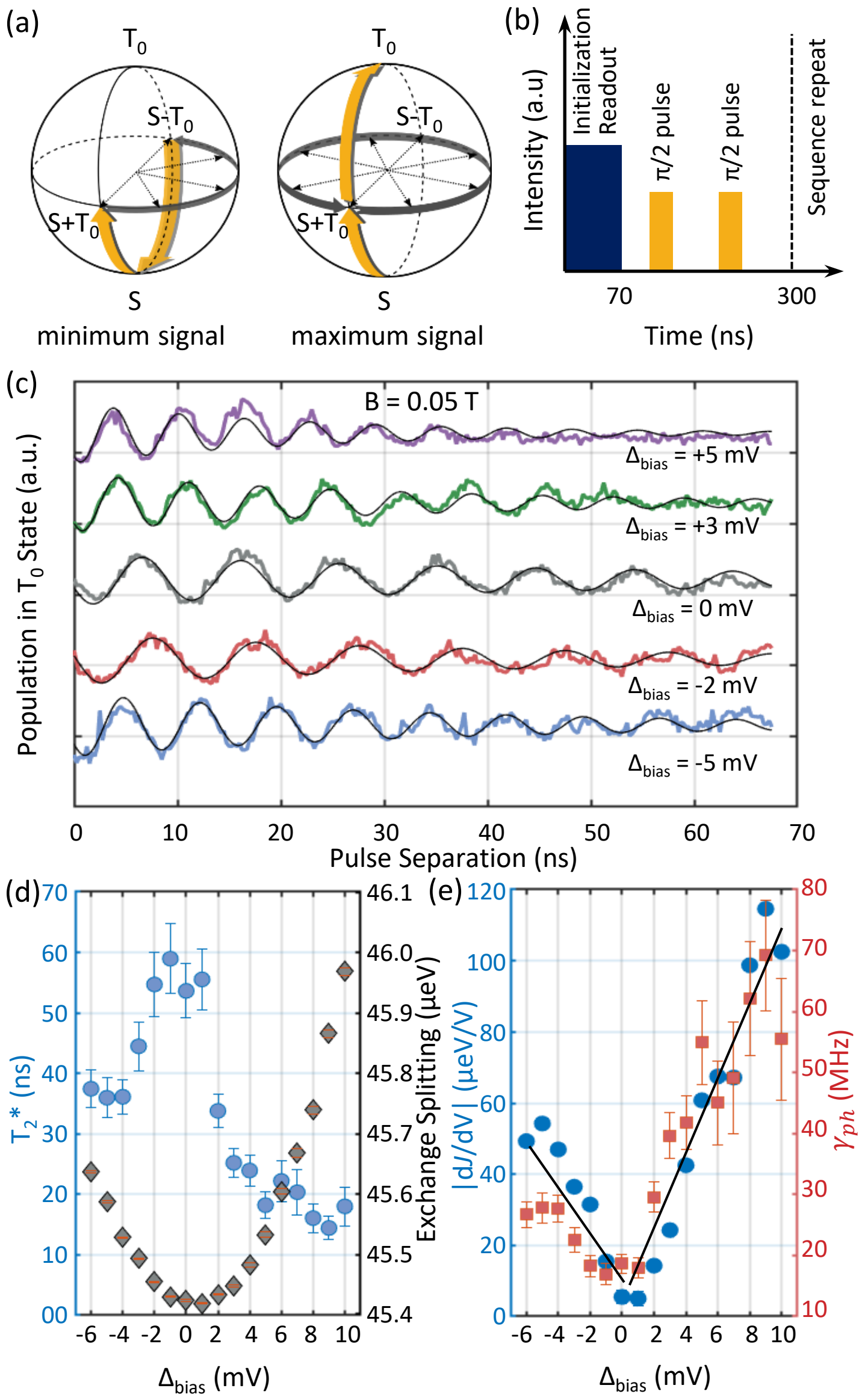

FIG. 3 (color online). (a) Bloch sphere representation of spin control of the $ST_0$ subspace. (b) The pulse sequence for Ramsey measurement which consists of a 70 ns initialization pulse, and two π/2 rotation pulses (2.5 ns duration) with variable separation between them. The sequence is repeated after every 300 ns. (c) Ramsey interference obtained at B = 0.05 T and at various biases $\Delta_{bias}$ relative to the sweet spot. The data is fitted with exponential oscillation decay (black solid lines). (d) $T_2^*$ (left vertical axis) and S-$T_0$ energy splitting (right vertical axis) are extracted from the Ramsey interference data and are plotted as functions of the bias offset from the sweet spot. (e) The dephasing rate ($\gamma_{ph}$) and magnitude of the derivative of the exchange splitting with respect to bias ($|dJ/dV|$) are plotted against the bias offset from the sweet spot showing a linear correlation between the two quantities.

To measure the coherence time $T_2^*$, we employ the Ramsey interferometry technique which is schematically shown in Fig. 3(a-b). The spin is initialized to the S state (south pole). The first π/2 rotation pulse (2.5 ns) creates the coherent superposition of S and $T_0$ states which then precesses and decays in a characteristic time scale $T_2^*$. A second π/2 rotation pulse converts the coherence into a population difference, which is read out by the initialization/readout pulse. As the separation between the two rotation pulses increases, the phase difference between the superposition and the second π/2 pulse increases, leading to a final state oscillating between $T_0$ (maximum signal) and $S$(minimum signal). Fig. 3(c) shows the Ramsey interference curves for five different bias offsets $\Delta_{bias}$ from the sweet spot at a relatively low magnetic field 0.05 T. In these measurements, we deliberately red-detune the microwave modulation frequency away from the $ST_0$ energy splitting by a few hundred megahertz. The detuning δ gives rise to oscillations in the Ramsey signal that are fit to precisely determine δ. The exchange splitting $J$ is then computed from δ and the microwave modulation frequency $\omega_m$ by the formula: $J = 2\omega_m + \delta$. On the other hand, $T_2^*$ is extracted from the exponential decay of the fringe contrast. $T_2^*$ and $J$ are plotted as a function of $\Delta_{bias}$ in Fig. 3(d).

We see a dramatic enhancement of $T_2^*$ in the vicinity of the sweet spot [Fig. 3(d)] with the maximum value of almost 60 ns. This value is comparable to that of the electrostatically confined GaAs QDM [9,23] taken at much lower temperature. An important question remains: what factors limit the coherence in our sample? As the spin relaxation time $T_1$ exceeds 1.5 μs in the optical pumping bias range, $T_1$ decay is not an important source of decoherence [28]. As mentioned previously, dephasing will occur from fluctuations in the bias voltage $\Delta V_{noise}$ (from the source or the local environment) that lead to fluctuations in $J$ [25,37,38]. To first order in the bias fluctuations, the dephasing rate is given by $\gamma_{\mathrm{elec}} = (dJ/dV)\,\Delta V_{noise}$. Indeed, we see a strong linear correlation between the dephasing rate denoted $\gamma_{ph} = 1/T_2^*$ and the derivative of $J$ with respect to the bias voltage V as shown in Fig. 3(e). From fitting this data, we extract $\Delta V_{noise} = 1.836 \pm 0.0262$ mV, which signifies the voltage fluctuation in our QDM. A bias-independent offset is included in the fit, with a value of $14.83 \pm 3.732$ MHz, to account for other sources of dephasing. This linear correlation strongly suggests that voltage noise is the primary dephasing mechanism away from the sweet spot [23,39]. This noise is expected to have a low frequency spectrum compared to the dephasing rate. We also attempt to partially bypass this noise source by apply a spin-echo pulse sequence and indeed measure much longer $T_2$, up to 500 ns as shown in the supplementary [28].

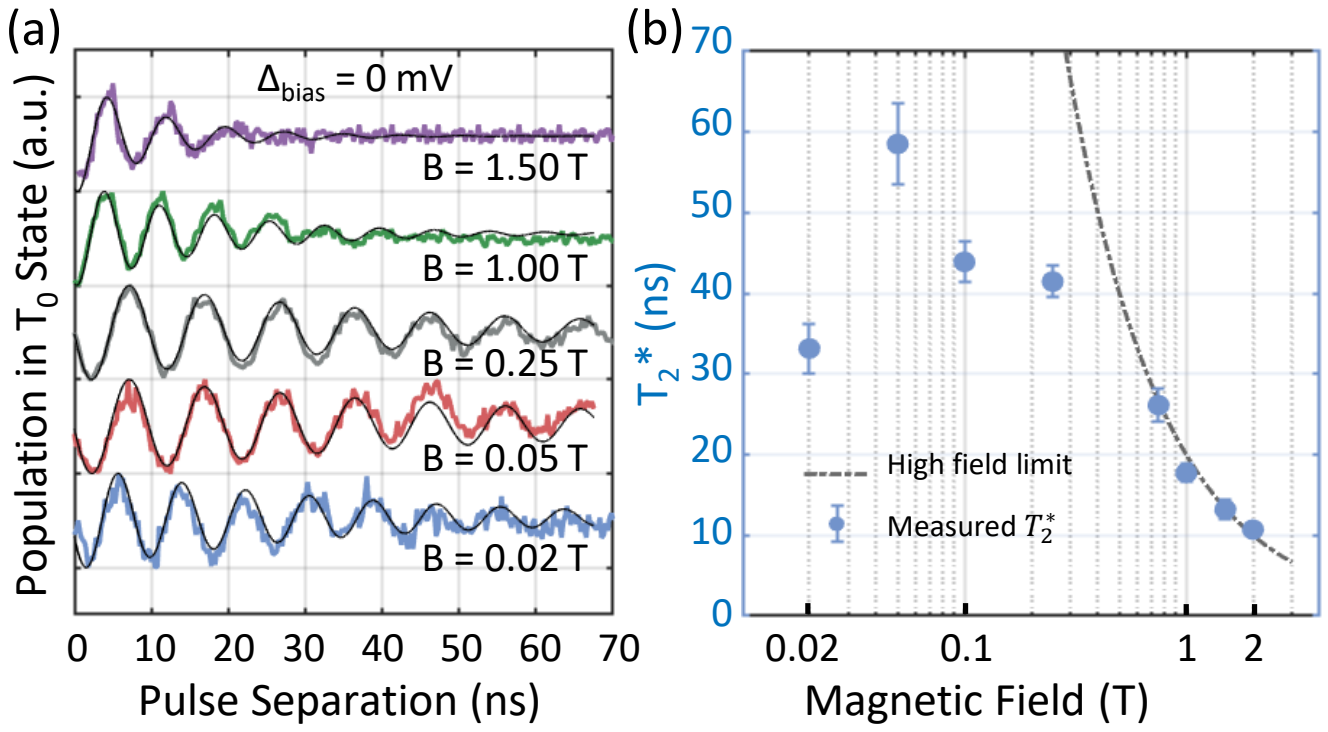

FIG. 4 (color online). (a) Ramsey interference data obtained at the sweet spot at various magnetic field strength. (b) Semilog plot of $T_2^*$ against the magnetic field. Dotted curve denotes the expected change of $T_2^*$ at high field regime.

We suspect that the dephasing rate at the sweet spot is due to fluctuations in the nuclear spin environment. To investigate this possibility, we measure $T_2^*$ for a series of magnetic fields and plot the results at the sweet spot in Fig 4. The coherence time peaks at 0.05 T and decreases precipitously once the field deviates from it. At the low field regime, the nuclear polarization can exceed the Zeeman splitting between triplets ($T_0$, T+, T-) and thus, strongly mix them leading to rapid dephasing [22]. On the other hand, high fields cause mixing between S and $T_0$ due to an effective field gradient stemming from the difference in g-factor between two dots. It leads to the dependence of the exchange splitting on B which has been explored theoretically [26] and experimentally [40]. In fact, if the gradient field is much stronger than the exchange interaction, the eigenstates would become the product states $|\uparrow\downarrow\rangle$ and $|\downarrow\uparrow\rangle$ of the two independent spins, in which we would exactly recover the dephasing of the single spin. The exchange splitting of the new eigenstates is modified by: $J' = \sqrt{J^2 + 4(\Delta g \mu_B B)^2}$. Here, $J$ is the exchange splitting at zero field; $\mu_B$ is the Bohr magneton; $\Delta g$ is the g-factor difference which can be measured directly by looking at the energy difference between $T_0 \leftrightarrow$R+ and T+$\leftrightarrow$R++ or $T_0 \leftrightarrow$R- and T-$\leftrightarrow$R-- at high magnetic field. In Fig. S6, we measure PL at 6 T and extract the average $\Delta g \mu_B$ of 0.3 $GHz/T$ which is about 5% of the electron spin g-factor in an isolated dot [41].

Similar to the case of bias fluctuations, we expect the dephasing rate from nuclear spin fluctuations in the high field regime to be given by $\gamma_{\text{nuc}} = \frac{dJ'}{d(\Delta g \mu_B B)} \Delta E_{nuc} \cong \frac{4\Delta g \mu_B B}{J} \Delta E_{nuc}$, where $\Delta E_{nuc}$ is the fluctuation in the difference in nuclear field energies between the two dots. The typical Overhauser field in QD is 20 mT [5–7] which gives $\Delta E_{nuc}$ of 120 MHz. We thus expect $\gamma_{\text{nuc}}/B$=13 MHz/T which is an order of magnitude in agreement with the measured dephasing rate at high fields i.e. $\gamma_{ph}/B \cong 50$ MHz/T at 2 T. This rough estimate of the expected dephasing contribution is plotted in Fig. 4(b), which matches quite well with the trend of the rapid dephasing at high fields after multiplying by a scaling factor. At moderate fields, the measured $T_2^*$ is much shorter than expected from this contribution, due to low field mixing of triplet states. The maximum $T_2^*$ occurs in an intermediate regime that balances dephasing from triplet mixing and dephasing from singlet-triplet mixing, due to $\Delta g$.

In summary, we present a systematic study of the coherence time $T_2^*$ in self-assembled QDM and demonstrate more than an order of magnitude improvement over electron spins in single QDs. We find that the electrical fluctuations mainly limit the spin coherence at biases away from the sweet spot. At the sweet spot, the nuclear spin fluctuations are largely responsible for dephasing in both low and high magnetic field regimes. Decreasing the impact of the nuclear spin fluctuations requires operating at magnetic fields larger than the nuclear fields that mix triplet states but not so large that differences in the g-factors give rise to singlet-triplet mixing. This singlet-triplet mixing can be avoided by reducing $\Delta g$, perhaps through growth of QDs with more similar thicknesses, and by increasing $J$. By reducing the tunnel barrier thickness or height, $J$ can be increased to several meV, but this must be balanced against the decreased spin relaxation times typically observed in these structures [22,39,42,43]. Improved spin coherence times for self-assembled QDs should enable a variety of applications in quantum photonics that require a quantum memory, such as photonic cluster state generation [44,45] and single photon switches and transistors [46].

Acknowledgements: We gratefully acknowledge helpful discussions with Edo Waks and Dima Farfurnik. This work was supported by the U.S. Office of Naval Research, the Defense Threat Reduction Agency (Grant No. HDTRA1-15-1-0011).

SUPPLEMENTAL MATERIAL

# Enhanced spin coherence at the sweet spot of a self-assembled quantum dot molecule

Kha X. Tran,[1] Allan S. Bracker,[2] Michael K. Yakes,[2†] Joel Q. Grim,[2], and Samuel G. Carter[2*]
[1] NRC Research Associate at the Naval Research Laboratory, 4555 Overlook Ave. SW, Washington, DC 20375, USA
[2] Naval Research Laboratory, 4555 Overlook Ave. SW, Washington, DC 20375, USA

[†]Present address: Air Force Office of Scientific Research, Arlington, VA 22203, USA.
*sam.carter@nrl.navy.mil

## I. Sample fabrication and characteristics:

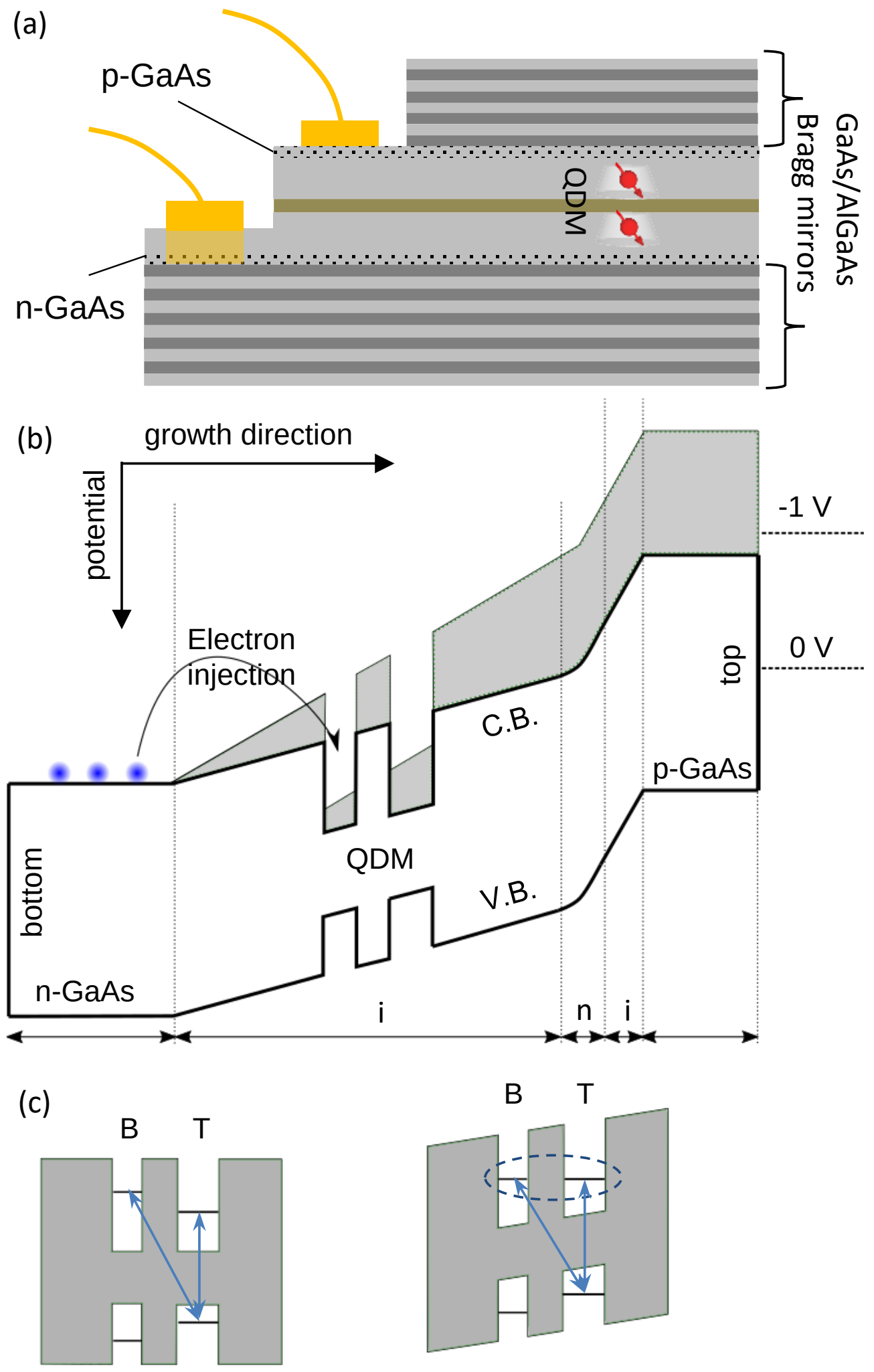

FIG. S1 (color online). (a) Schematic of the QDM sample structure. The electrodes are color coded yellow. (b) Band diagram of the device layer structure. (c) Flat band condition (left hand side) and forward bias condition (right hand side) where the electron tunneling between two dots occur. Here bottom (top) dot is labeled as B (T). The transitions are labeled as solid arrows.

The sample is grown by molecular beam epitaxy via Stransky-Krastanov method on an (100) n-doped GaAs substrate. The QDs in the second layer tend to nucleate directly on top of the QDs in the first layer due to the strain extension. The tunneling barrier between the top and bottom dots is 9 nm thick consisting of 3/2/4 nm GaAs/AlGaAs/GaAs layers. The whole structure is then embedded within the n-i-n-i-p diode structure and placed within a distributed Bragg reflector planar cavity to enhance the photon collection efficiency [Fig. S1(a)]. To charge the device, the gate voltage is applied between the bottom n-doped GaAs substrate and the p-type region. This allows the electrons to be controllably injected into the QDM. We further note that the dimension in Fig. S1(b) is not drawn to scale in which the QDM size is exaggerated for clarity.

The transition energy of the bottom dot is about tens of meV higher than that of the top dot. Under the forward bias condition [Fig. S1(c) right hand panel], the electron energy levels of the two dots are in resonance, allowing electron tunneling that forms bonding and anti-bonding states between the two dots. All of the transitions discussed in the main text are from the top dot.

The sample is mounted in a closed cycle, split-coil magneto-optical cryostat at ∼3K on three piezoelectric positioner stages. An aspheric lens is used for focusing excitation lasers onto the sample and for collecting emission. To locate the QDM with the sweet spot at the center of the two electron bias range, a photoluminescence (PL) vs bias map is taken programmatically on many locations on the sample using the positioner stages. To simplify the spectra, we take the PL at zero magnetic field. The QDMs are found based on the anti-crossing features [1,2] on the PL-bias map of the scan [Fig. S2(a)]. Higher spectral resolution PL capable of resolving the $ST_0$ splitting is taken again on those QDMs in order to determine the sweet spot bias and thus the most favorable QDM [Fig. S2(b)].

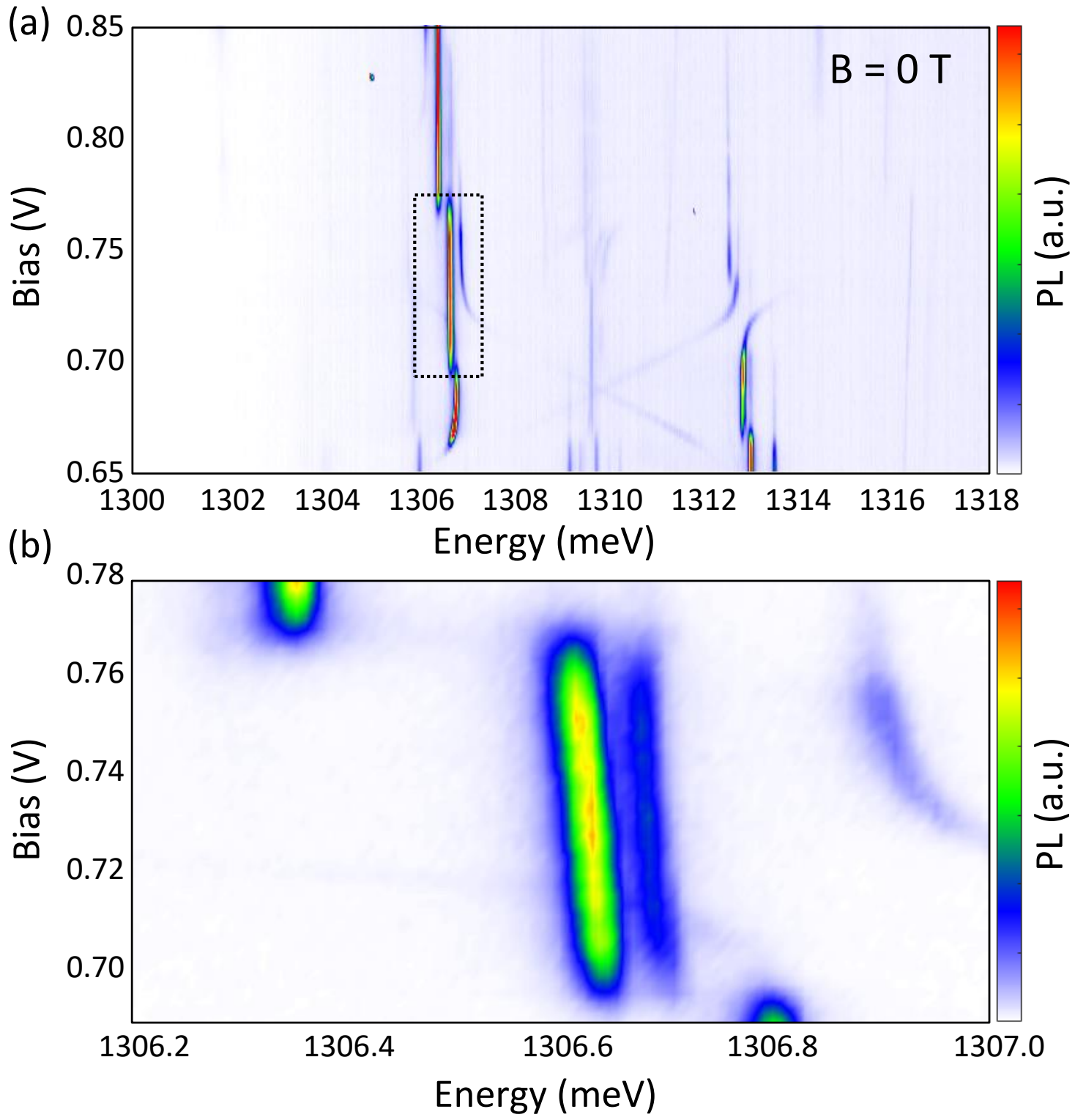

FIG. S2 (color online). (a) Low spectral resolution PL-bias map of the QDM studied in the main text showing anti-crossing resonances. (b) High spectral resolution PL-bias map of the same QDM is taken in the region marked by the dotted rectangular box in (a).

II. Spin relaxation ($T_1$) measurement

Fig. S3(a) upper panel depicts the pulse sequence for our $T_1$ measurement which consists of two 100 ns pulses separated by a variable delay time t. The pulses are resonant with $T_0$↔R- transition which gives rise to the Raman scattering S↔R- [Fig. S3(a) lower panel]. The Raman intensity from the readout pulse is generally smaller than that of the initialization pulse due to spin pumping. As the delay time increases, the readout Raman signal increases until it matches the Raman intensity from the first pulse when the spin completely relaxes between the S and $T_0$ states. By monitoring the readout Raman intensity versus delay time, we extract $T_1$ vs bias at 1T and plot it in Fig. S3(b).

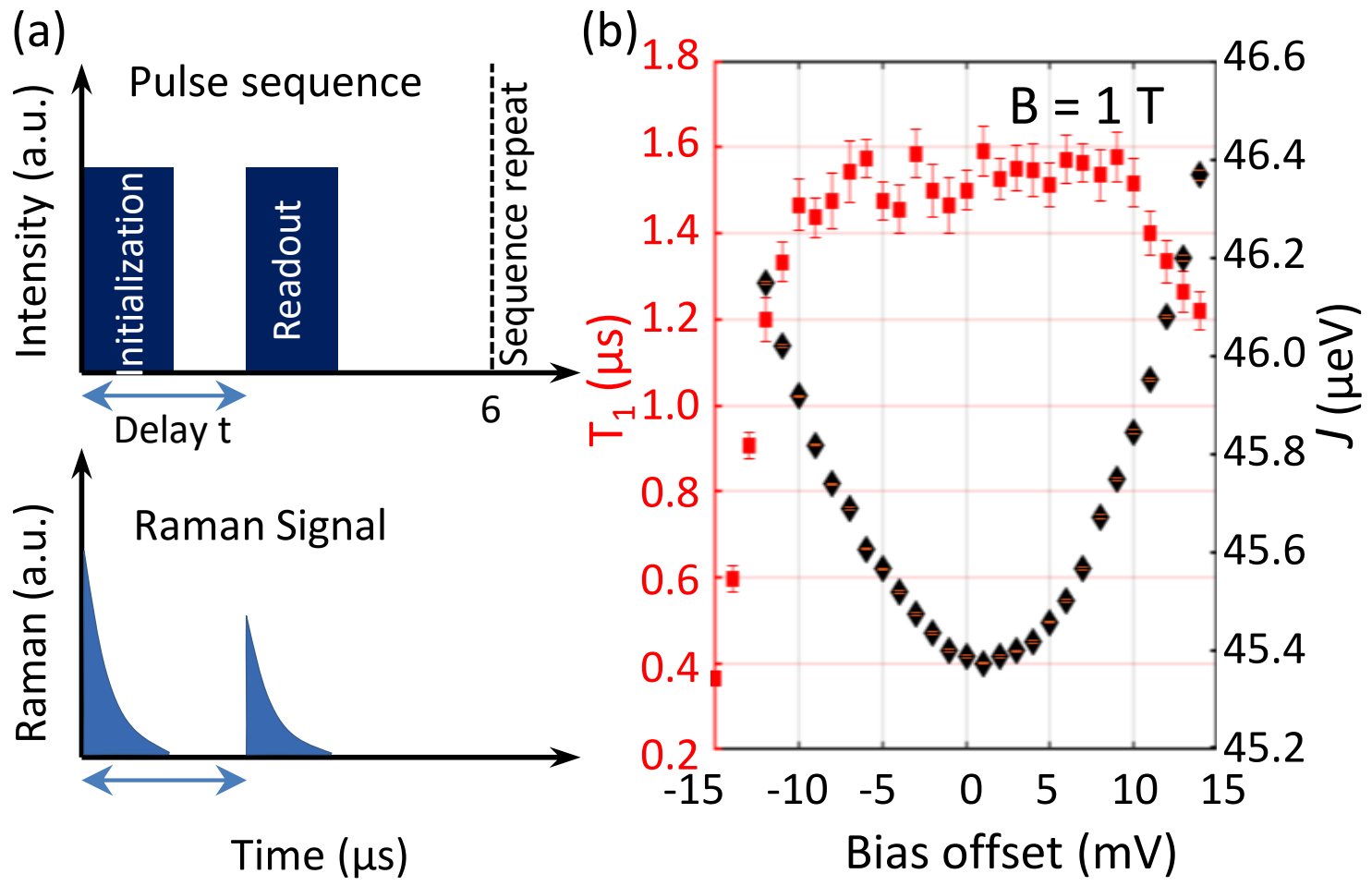

FIG. S3 (color online). (a) Top: schematic of the pulse sequence to measure $T_1$ which consists of two 100 ns initialization and readout pulses resonant with $T_0$↔R- transition. Bottom: the corresponding Raman signal from S↔R- transition. (b) $T_1$ as a functions of bias offset from the sweet spot at 1T magnetic field

III. Additional $T_2^*$ measurement vs magnetic field B

We show additional $T_2^*$ measurements in Fig. S4 for 0.25T, 1T, and 1.5T. As the field increases, the nuclear spin fluctuation effect dominates that of the electric noise. As a consequence, the enhancement of the coherence time at the sweet spot is less prominent (left side panel). The dephasing rate deviates from being proportional to $|dJ/dV|$ on the reverse bias side to some extent for reasons that are not yet clear (right side panel).

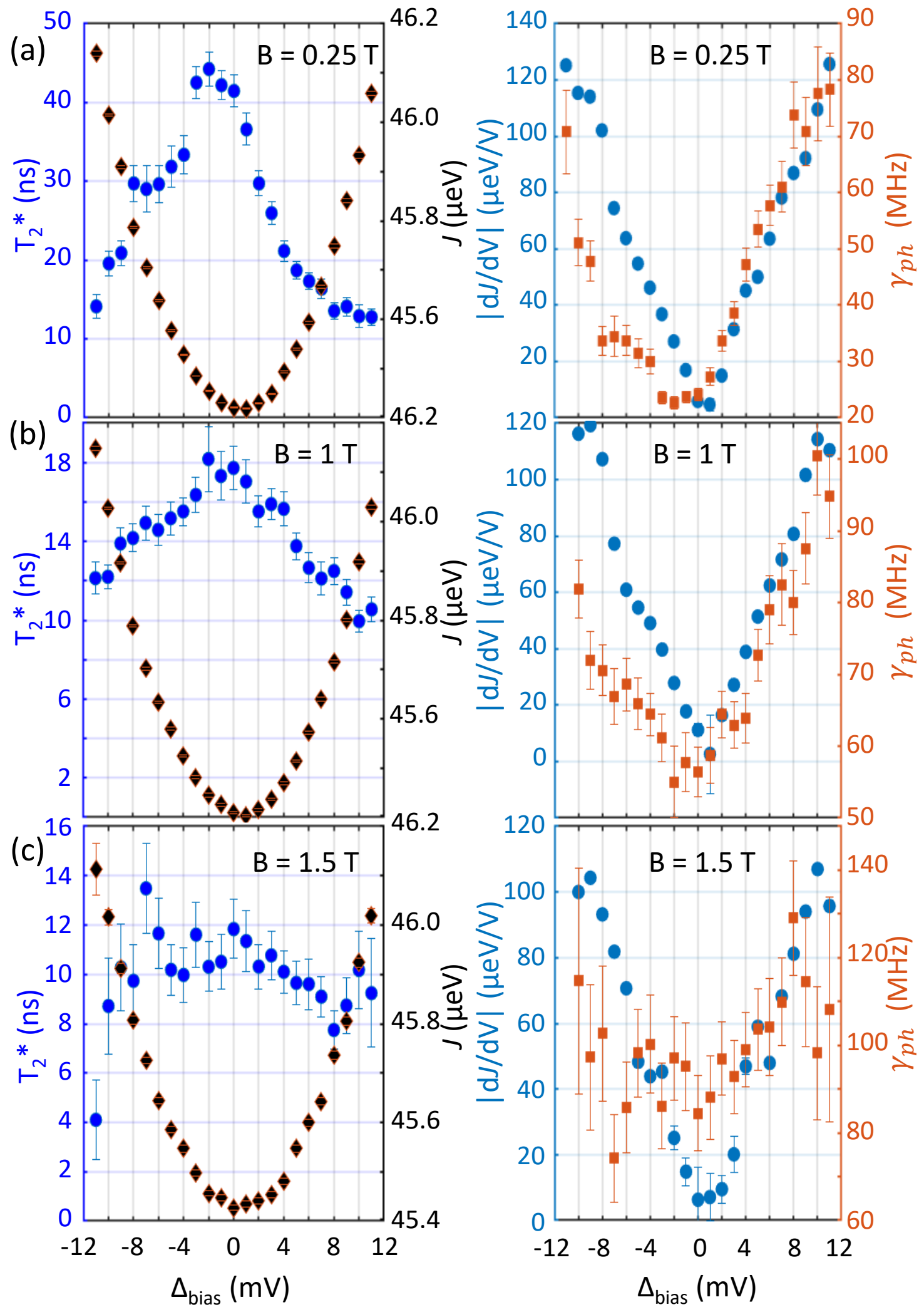

FIG. S4 (color online). Left hand side: $T_2^*$ and *J* vs bias offset from the sweet spot. Right hand side: the corresponding d*J*/dV and the dephasing rate $\gamma_{ph} = 1/T_2^*$ vs bias offset for different magnetic field strength (a) 0.25T; (b) 1T; (c) 1.5T

## IV. $T_2$ measurement vs magnetic field B

The pulse sequence for spin-echo measurement is shown in Fig. S5(a). As the separation time t between two π/2 pulses increases, the π pulse is always in the middle (t/2) of these two pulses. Spin-echo measurement removes the effects of low frequency noise leaving the faster fluctuations [3,4]. As a result, $T_2$ is significantly extended up to 500 ns at the sweet spot at 0.05 T. This value is still less than the $T_2$ for single dot [3] which could be extended up to microseconds. We suspect that phonon-coupling is another source of dephasing in QDM as suggested by the theoretically studies on the gate defined QDMs [5,6]. We would also expect the electric field fluctuations to have low frequency spectrums that would largely be eliminated by the spin echo technique, but there is still a clear peak at the sweet spot. Further study would be needed to quantify this dephasing mechanism.

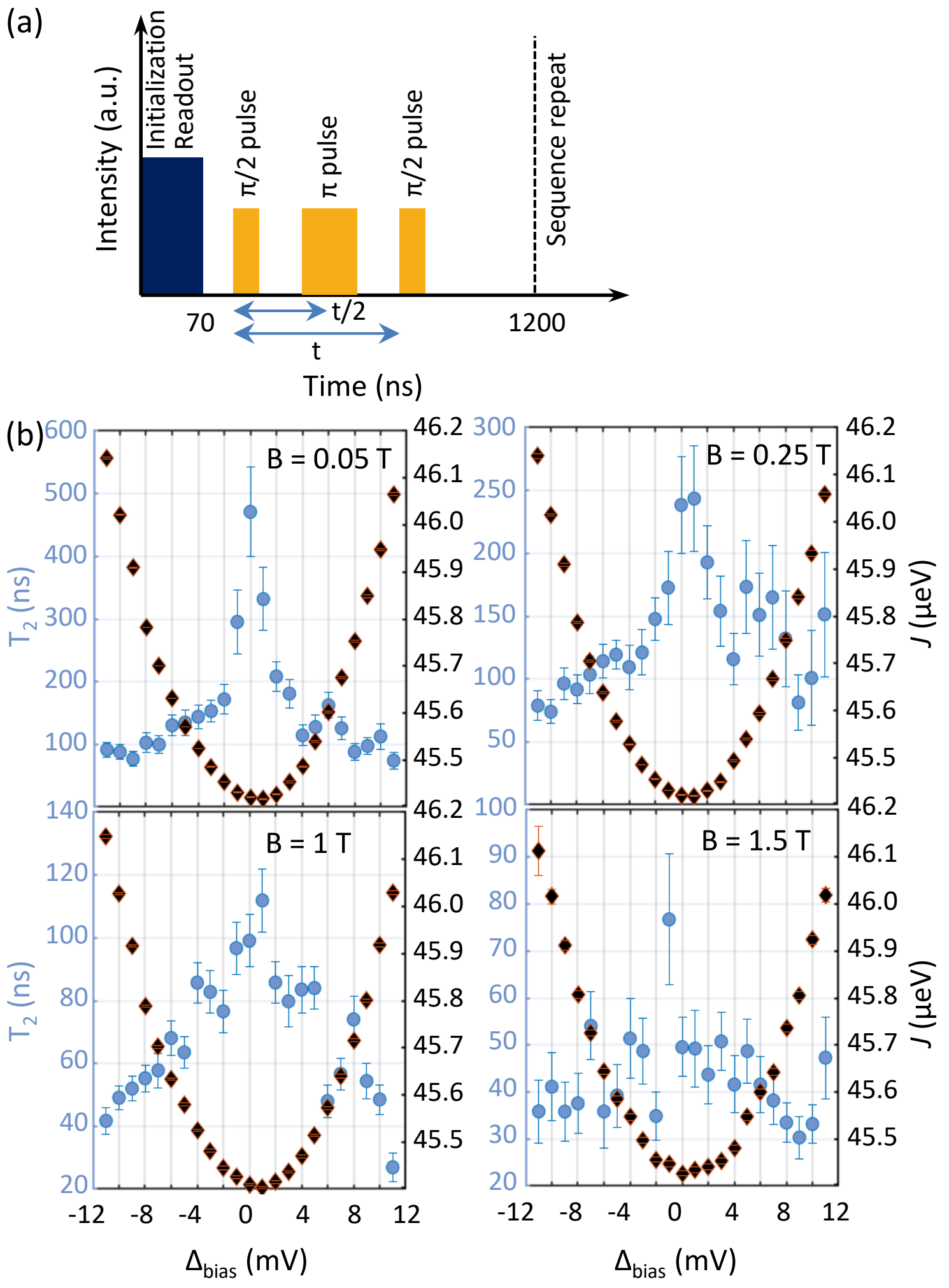

FIG. S5 (color online). (a) Schematic of the pulse sequence for the spin echo technique. (b) $T_2$ and $J$ vs bias offset for different magnetic field strengths

## V. g-factor difference of the two dots

Fig. S6(a) shows the PL spectra of the QDM at a few bias offsets from the sweet spot at 6 T. Extremely high spectral resolution $\sim 2\ \mu eV$ is achieved by passing the PL signal through a fiber Fabry Perot interferometer before being detected by a spectrometer. The difference in g-factor between the two dots breaks the energy degeneracy between the lower transition i.e. T-↔R-- and $T_0$↔R- and between the higher transitions i.e. T+↔R++ and $T_0$↔R+ which are shaded red and blue [Fig. S6(a)] respectively. We notice that the lower and higher transition splittings give different values of $\Delta g\mu_B$ [Fig. S6(b)]; however, they are within the uncertainty $\sim 2\ \mu eV$ of our measurement. Thus, we take the average of all the extracted splittings and quote $\Delta g\mu_B \sim 0.3\ GHz/T$ in the main text.

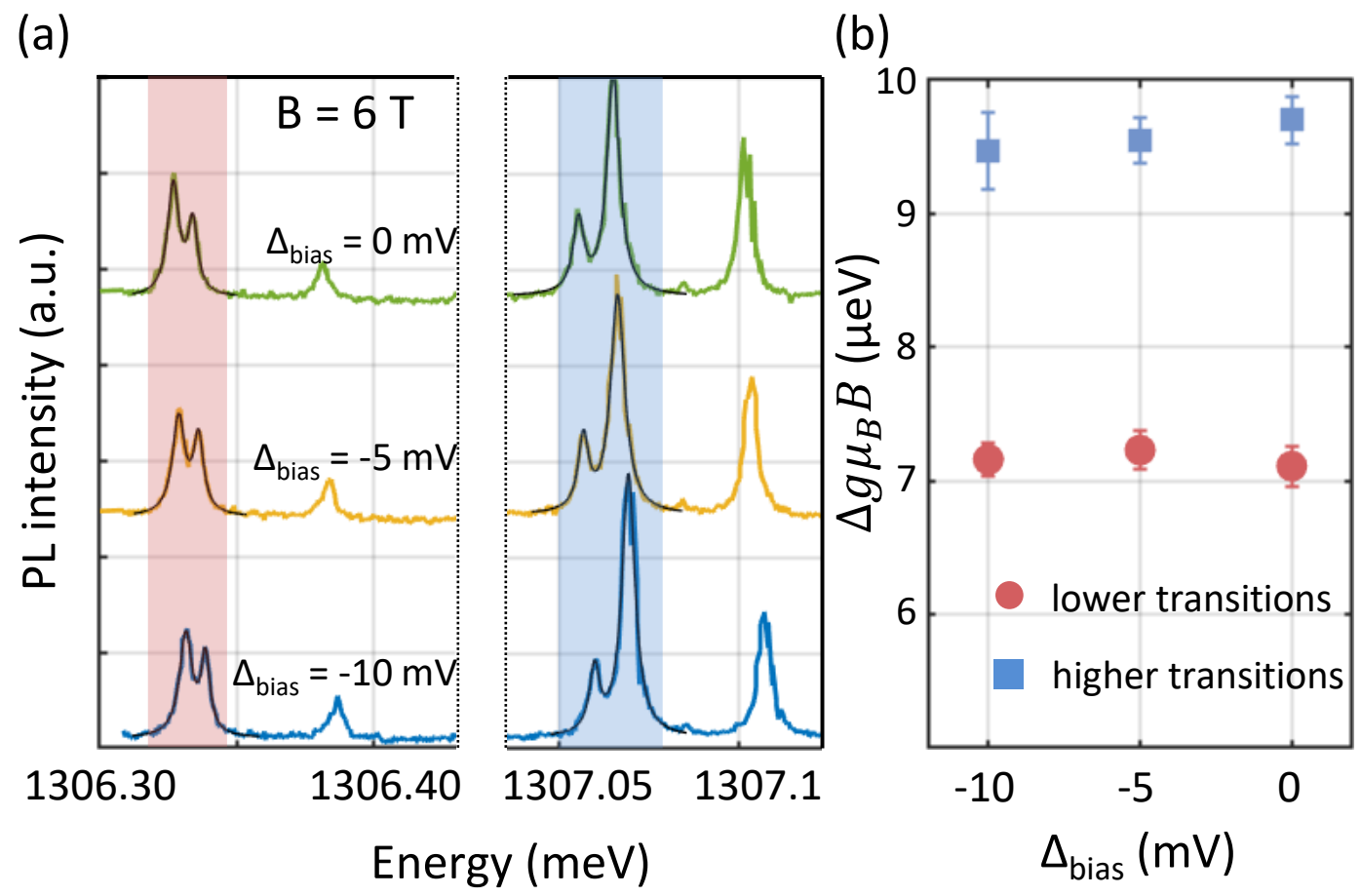

FIG. S6 (color online). (a) PL measurements at 6 T at a few different biases (b) Energy splittings extracted from (a) between T-↔R-- and $T_0$↔R- (lower transitions); T+↔R++ and $T_0$↔R+ (higher transitions) yield the measurement of $\Delta g \mu_B$